\documentclass[a4paper, 12pt]{article}
\usepackage{eurosym}
\usepackage[a4paper,margin=1in]{geometry}
\usepackage[round,sort]{natbib}
\usepackage[english]{babel}
\usepackage{numprint}
\usepackage{float}
\usepackage{graphicx}
\usepackage{dcolumn}
\usepackage{csquotes}
\usepackage{amsmath}
\usepackage{caption}
\usepackage[normalem]{ulem}
\usepackage[dvipsnames]{xcolor}
\usepackage{subcaption}
\usepackage{caption}
\npthousandsep{,}

\begin{document}

\title{\textbf{A Study Protocol for an Instrumental Variables Analysis of the Comparative Effectiveness of two Prostate Cancer Drugs}}
\author{Per Johansson \hspace{.2cm} \\
%EndAName
Department of Statistics, Uppsala University, IFAU and Tsinghua University\\
Paulina Jon\'eus \\
Department of Statistics, Uppsala University\\
Sophie Langenskiold,  Department of Medical Sciences, Cardiology}
\maketitle

\begin{abstract}
This paper presents a protocol, or design, for the analysis of a comparative effectiveness evaluation of abiraterone acetate against enzalutamide, two drugs given to prostate cancer patients. The design explicitly make use of differences in
prescription practices across 21 Swedish county councils for the estimation of the two drugs comparative effectiveness on overall mortality, pain and skeleton related events. The design requires that the county factor: (1) affects the probability to be treated (i.e. being prescribed abiraterone acetate instead of enzalutamide) but (2) is not otherwise correlated with the outcome. The fist assumption is validated in the data.  The latter assumption may be untenable and also not possible to formally test. However, the validity of this assumption is evaluated in a sensitivity analysis, where data on the two morbidity outcomes (i.e. pain and skeleton related events) observed before prescription date are used. We find that the county factor does \emph{not} explain these two pre-measured outcomes. The implication is that we cannot reject the validity of the design. 
\end{abstract}

{\ }

\bigskip

\vfill \newpage

\section{Introduction}

The aim of this paper is to present a study protocol, or design, for a comparative effectiveness analysis of abiraterone acetate (AA) against enzalutamide (ENZ) for prostate cancer (PC) patients on overall mortality, pain and skeleton related events. PC is the second most commonly diagnosed solid organ malignancy in the world and it is the second leading cause of cancer death. The mortality arise when the patients have progressed to the advanced stage, metastatic castrate resistant prostate cancer (mCRPC). Novel and expensive antiandrogenic medications (NAM) are suggested for treatment in mCRPC patients. AA and ENZ are two types of these NAM. Thus, the results from later analysis should be of general interest.

The design makes use of differences in prescription practices across 21 county councils. The design is known as an Instrumental Variables (IV) design, where the county factor is the IV. It can be seen as a `natural experiment', that is we have IV that: (1) affect the probability of treatment (i.e. AA or ENZ) at the individual level, but (2) is not otherwise correlated with the outcome.\footnote{In the medical literature (cf. \cite{Cawley_2015,Brookhart_Schneeweiss_2007,Brookhart_etal_2006}) this type of IV are sometimes denoted ‘preference-based’ IV. The definitions is best suitable when making use of doctors differences in preferences. Due to social interaction individual preferences may lead to `preference differences' across hospitals and also across county councils.} 

One relevant concern with the design is that population health and health care quality differs across the 21 Swedish county councils. If this is the case, the county factor IV is correlated with the outcome and this invalidate the design. For this reason detailed data on observed health- and socioeconomic status of the patients is included in the design. In addition, mortality data in prostate cancer at the county level is included in the design. Requirement (1) is supported in data as the county factor is shown to sufficiently shift the individual probability to be prescribed AA or ENZ, given the observed covariates. Adding covariates is however not sufficient for the design to be valid, that is requirement (2) may not hold. 

Results from conditional (on covariates) IV-analyses is less credible than the results from unconditional IV-analyses. The first concern with the conditional IV-analysis is that it is in general not know which covariates to add to make the design valid. The second concern is that the researchers using IV-analyses observing the outcome may conflate the design with analysis. Thus, there is a risk of choosing the model specification (by e.g. adding or removing covariates) to obtain statistical significant result. By publishing the IV-design before observing the outcomes this second form of bias is controlled for. According to our understanding, this is the first study that publish an IV-design in a pre-analysis plan and this even before observing our three outcomes after the date of prescription. However, also the first concern is addressed in this protocol. We do this by providing as a sensitivity analysis of the validity (2) of the IV-design by using data on the two morbidity outcomes (pain and skeleton related events) observed before  prescription date. The sensitivity analysis is done by estimating the comparative effect of the county factor on these outcomes on data in the period between diagnosis and prescription. If the county factor does not contribute in explaining these two pre-measured outcomes, the validity of the design is not rejected. The resulting test showed that we could not reject the null of no effects from the county factor on these outcomes.

The paper makes use of data from the study protocol of \cite{Johansson_et_al_2021a} for the same comparative effectiveness evaluation. Their design assumes observed covariates on health and socioeconomic status controls for all potential confounding.  A comparison of the results from the two studies will provide an understanding of pros and cons of the two types of design. This might stimulate a methodological discussion in the field of comparative effectiveness research. As both designs could be valid, would similar results strengthen the causal interpretation of the results.

The rest of the paper has the following structure. Section \ref{sec:Data} describes the data. Section \ref{sec:idIV} presents the instrumental variables design. The final evaluation is given in Section \ref{sec:Plan}. The paper
ends with a discussion in Section \ref{sec:Discussion}.

\section{Data}

\label{sec:Data}

Data in \cite{Johansson_et_al_2021a} was collected from population registers
administrated by the National Board of Health and Welfare (NBHW), and
Statistics Sweden (SCB).  The population was then restricted to all men
collecting a prescription of AA or ENZ during the period June 1 2015
to June 15 2018, in total \numprint{4601} men. For this population the year
of the diagnosis ranges between 1986 and 2016. Consequently, there is
substantial variation in the time to be prescribed AA or ENZ from the
date of diagnosis. 

As can be seen from Table \ref{Regdiff} the prescription of the two drugs
varies over the 21 county councils, hereafter denoted counties, the
responsible body for health care in Sweden. The fact that the prescription
varies over counties is a notable finding as it suggests differences in
prescription that may not be related only to the health status of the patients. From this table we can see that in total, $24\%$ of the patients were
prescribed AA.

\begin{table}[]
\caption{Proportion prescribed ENZ and AA respectively, per county
and in total}
\label{Regdiff}\centering
{\footnotesize 
\centerline{
\begin{tabular}{lrrrrrrrrrrcc}
\\[-1.8ex]
\hline
&\multicolumn{4}{c}{ENZ} && \multicolumn{4}{c}{AA } && ENZ, & AA, \\ 
&2015&2016&2017&2018& & 2015&2016&2017&2018&&total & total \\ 
\hline
\\[-1.8ex]
Blekinge & 0.83 & 0.92 & 0.84 & 0.80 & \vline & 0.17 & 0.08 & 0.16 & 0.20 & \vline & 0.85 & 0.15 \\ 
Dalarna & 0.86 & 0.65 & 0.54 & 0.62 & \vline & 0.14 & 0.35 & 0.46 & 0.38 & \vline & 0.69 & 0.31 \\ 
Gavleborg & 0.60 & 0.92 & 0.93 & 0.96 & \vline & 0.40 & 0.08 & 0.07 & 0.04 & \vline & 0.85 & 0.15 \\ 
Gotland & 0.73 & 0.92 & 1.00 & 1.00 & \vline & 0.27 & 0.08 & & & \vline & 0.87 & 0.13 \\ 
Halland & 0.71 & 0.89 & 0.89 & 0.73 & \vline & 0.29 & 0.11 & 0.11 & 0.27 & \vline & 0.81 & 0.19 \\ 
Jamtland & 0.50 & 0.81 & 0.76 & 0.75 & \vline & 0.50 & 0.19 & 0.24 & 0.25 & \vline & 0.71 & 0.29 \\ 
Jonkopings lan & 0.65 & 0.94 & 0.89 & 0.76 & \vline & 0.35 & 0.06 & 0.11 & 0.24 & \vline & 0.82 & 0.18 \\ 
Kalmar & 0.72 & 0.88 & 0.85 & 0.93 & \vline & 0.28 & 0.12 & 0.15 & 0.07 & \vline & 0.84 & 0.16 \\ 
Kronoberg & 0.49 & 0.43 & 0.21 & 0.33 & \vline & 0.51 & 0.57 & 0.79 & 0.67 & \vline & 0.39 & 0.61 \\ 
Norrbotten & 0.77 & 0.85 & 0.64 & 0.65 & \vline & 0.23 & 0.15 & 0.36 & 0.35 & \vline & 0.72 & 0.28 \\ 
Orebro & 0.75 & 0.96 & 0.68 & 0.43 & \vline & 0.25 & 0.04 & 0.32 & 0.57 & \vline & 0.74 & 0.26 \\ 
Ostergotlands lan & 0.36 & 0.91 & 0.86 & 0.90 & \vline & 0.64 & 0.09 & 0.14 & 0.10 & \vline & 0.82 & 0.18 \\ 
Skane & 0.88 & 0.96 & 0.95 & 0.83 & \vline & 0.12 & 0.04 & 0.05 & 0.17 & \vline & 0.92 & 0.08 \\ 
Sodermanland & 0.80 & 1.00 & 1.00 & 0.88 & \vline & 0.20 & & & 0.12 & \vline & 0.92 & 0.08 \\ 
Stockholm & 0.70 & 0.82 & 0.80 & 0.79 & \vline & 0.30 & 0.18 & 0.20 & 0.21 & \vline & 0.78 & 0.22 \\ 
Uppsala & 0.66 & 0.57 & 0.26 & 0.38 & \vline & 0.34 & 0.43 & 0.74 & 0.62 & \vline & 0.49 & 0.51 \\ 
Varmland & 0.87 & 0.96 & 0.67 & 0.75 & \vline & 0.13 & 0.04 & 0.33 & 0.25 & \vline & 0.83 & 0.17 \\ 
Vasterbotten & 0.64 & 0.71 & 0.64 & 0.76 & \vline & 0.36 & 0.29 & 0.36 & 0.24 & \vline & 0.68 & 0.32 \\ 
Vasternorrland & 0.47 & 0.71 & 0.74 & 0.85 & \vline & 0.53 & 0.29 & 0.26 & 0.15 & \vline & 0.64 & 0.36 \\ 
Vastmanland & 0.74 & 0.74 & 0.69 & 0.64 & \vline & 0.26 & 0.26 & 0.31 & 0.36 & \vline & 0.71 & 0.29 \\ 
Vastra gotalands lan & 0.50 & 0.78 & 0.68 & 0.65 & \vline & 0.50 & 0.22 & 0.32 & 0.35 & \vline & 0.66 & 0.34 \\ 
\\[-1.8ex]
\hline
\\[-1.8ex]
Total & 0.68 & 0.83 & 0.76 & 0.74 & & 0.32 & 0.17 & 0.24 & 0.26 & & 0.76 & 0.24 \\ 
\hline
\end{tabular}
} }
\end{table}

All inpatient care visits in Sweden and all prescribed drugs are registered in the registers from NBHW. The inpatient care register contains
among others information on all diagnoses (using the ICD classifications), the date of admission and discharge. The pharmaceutical register contains the date of prescribing and dispensing of drugs, and also, the ATC class of the drug. We derive, among others, the number of health care visits and we include covariates that measure the health status before the prescription of the two treatments.

We construct 23 continuous covariates measuring the general health and health progression of the patient, both before diagnosis and between diagnosis and treatment, including number of visits at different periods and number of days in inpatient care. The general health status is in addition captured by the inclusion of the Elixhouser comorbidity index at diagnosis. We also include covariates separately for diseases deemed most important, that is, cardiovascular diseases, metastases, diabetes, fatigue and osteoporosis (see Table \ref{diseases} in Appendix for the included ICD codes). This results in 16 continuous covariates on number of visits and 8 indicators on whether a patient have had the diagnosis or not. We also derive 3 covariates measuring the number of collected prescriptions on medications, three years before the treatment, related to cardiovascular diseases and diabetes.

From the SCB data we create 91 variables that supposedly will be able to describe the socio-economic status of the patient three years before the diagnosis and three years before treatment respectively, with data from 1991 until 2015 \footnote{%
For patients with diagnoses before or after these years, information on
socioeconomic status is given by values from a year as close as the
diagnosis as possible.} This includes information on age, marital status and
educational level as well as pensions, income, sick leave and other security
benefits for the patient and the household. In the few cases of partly missing values on continuous covariates, the mean over the three preceding years is used.

Educational level is the
highest completed education and is classified as less than, equal to or more
than secondary school. There were 33 observations where education was
reported being unknown. Here a $5$ nearest neighbour approach is used to
impute the missing values. That is, the most common value of the five
patients, i.e., neighbours, who are most similar when it comes to income,
pension, age and country of birth, is imputed for every missing value of the
categorical variable measuring educational level.

A potential problem is that we do not observe whether the patients have received chemotherapy or not in our data. It is however important to note that we observe and include the time between diagnosis and treatment as a control variable. In addition, as the quality of the health care affects the health and as it may be related to the prescription of the two drugs, we also include the historical county specific mortality related to prostate cancer at the year of diagnosis.

All 144 covariates with descriptions are
presented in Table \ref{All_vars} in Appendix. Table \ref{sum_stat_intro} provides summary statistics of the AA and ENZ patients for a subset of 15 variables deemed to be the most important. From this table we can see that the two groups in general are very similar, there are for instance no significant differences in average age, educational level or marital status.

\begin{table}[]
\caption{Summary statistics}
\label{sum_stat_intro}
\centering
{\footnotesize 
\centerline{
\begin{tabular}{lrrrrr}
\hline 
\\[-1.8ex]
Description & ENZ & AA & Difference \\ 
\hline
\\[-1.8ex]
Age at treatment & 75.27 (7.85) & 75.29 (7.70) & -0.02\\ 
Years to treatment from diagnosis & 6.95 (5.00) & 7.29 (5.31) & -0.34$^{*}$\\ 
Less than secondary school education & 0.36 (0.48) & 0.34 (0.48) & 0.02 \\ 
Secondary school education & 0.39 (0.49) & 0.39 (0.49) & 0.00\\ 
Living with a partner & 0.66 (0.47) & 0.67 (0.47) & -0.01 \\ 
Country specific mortality at diagnosis, deaths per $1000$ inhabitants & 0.05 (0.01) & 0.05 (0.01) & 0.00  \\
Drugs used in diabetes, ATC A10, 3 years before treatment & 0.16 (0.37) & 0.12 (0.33) & 0.04$^{***}$ \\ 
Beta blocking agents, ATC C07, 3 years before treatment & 0.38 (0.49) & 0.37 (0.48) & 0.01 \\ 
Calcium channel blockers, ATC C08, 3 years before treatment& 0.31 (0.46) & 0.32 (0.47) & -0.01 \\ 
Elixhouser score at diagnosis $=1-4$ & 0.39 (0.49) & 0.37 (0.48) & 0.02 \\ 
Elixhouser score at diagnosis $>=5$ & 0.02 (0.15) & 0.02 (0.15) & 0.00 \\ 
Osteoporosis before treatment& 0.01 (0.11) & 0.01 (0.10) & 0.00 \\ 
Metastases before treatment& 0.70 (0.46) & 0.75 (0.43) & -0.05$^{***}$ \\ 
Acute myocardial infarction before treatment (I21)& 0.10 (0.30) & 0.07 (0.25) & 0.03$^{***}$ \\ 
Atrial fibrillation and flutter before treatment (I48) & 0.16 (0.36) & 0.18 (0.38) & -0.02 \\ 
Other cardiovascular diseases before treatment & 0.27 (0.45) & 0.26 (0.44) & 0.02 \\ 
Fatigue before treatment & 0.05 (0.22) & 0.04 (0.20) & 0.01 \\ 
\\[-1.8ex] 
\hline 
\multicolumn{3}{l}{Standard deviations within parentheses. $^{*}p<$0.1; $^{**}p<$0.05; $^{***}p<$0.01.} \\ 
\end{tabular}} }
\end{table}

The average time in years to treatment from diagnosis is somewhat longer for patient prescribed AA. ENZ patients have a higher
prevalence of acute myocardial infarction, a result that supports the conclusion from the qualitative study described in \cite{Johansson_et_al_2021a}, while there are no significant
differences when it comes to atrial fibrillation and flutter. Also in agreement with the qualitative study, we find a higher prevalence of prescriptions of drugs used in diabetes among ENZ patients three years before treatment. In addition, the prevalence of metastases at the time of treatment is higher among AA patients.

\section{Instrumental Variable Analysis}
\label{sec:idIV}

The design is based on the Neyman-Rubin potential outcomes framework
\citep{Neyman1923, Rubin1974, Holland1986}. Define the potential outcome if
a patient has been given AA $Y(1)$ and $Y(0)$ if he instead would have been given
ENZ, and let $D=1$ if the patient was prescribed AA and $D=0$ if
prescribed ENZ. We observe the outcome $Y_{i},$ the treatment, $D_{i}$
and a large set of covariates, $\mathbf{X}_{i}$ for a sample or population
of $n$ individuals, that is $i=1,...n$.

For the time being, assume that doctors in randomly chosen counties were
encouraged (e.g. through economic incentives or relevant information on the
drugs) to prescribe AA over ENZ, while doctors in other counties were
not. Let $Q=1$ if a patient meets with a doctor who were encouraged to
prescribe AA, while $Q=0$ if the patient meets a doctor who did not get
the encouragement. As the prescription decision is up to the doctor and
potentially the patient it is not likely that the doctors encouraged to
prescribe AA always prescribe AA. However, it is likely that the
encouragement increases the likelihood of prescribing AA, which then can
be used in the estimation of the complier effect, that is, the effect on the
patients that meet with a doctor that prescribed AA as consequence of
the encouragement.

To formalize the identification strategy, let $D(1)$ and $D(0)$ be the
potential drug prescription meeting a doctor encouraged to subscribe AA
and not encouraged, respectively. The observed drug prescription is then
equal to

\begin{equation}
D=(1-Q)D(0)+QD(1)
\end{equation}

In addition to the SUTVA, assume (i) $Q$ $\perp \! \! \! \perp
(D(0),D(1),Y(1),Y(0)),$ (ii) $P(D=1|Q=1)\neq P(D=1|Q=0)$ and (iii) $D(1)\geq
D(0)$ (no \textit{defiers}, or monotonicity). These assumptions allow us to
identify the complier treatment effect or the local average treatment effect
(LATE). 

In this hypothetical \enquote*{encouragement} experiment it would be likely
that all these three assumptions are valid: (i) hold due to the
randomization, (ii) holds if the encouragement is seen as useful for the
doctors and assumption (iii) is also valid if the patient believes the
doctors have better understanding of the drugs than them. That is, a patient
who meets with a doctor encouraged to prescribe AA would not be more
likely to refuse, and thus demand ENZ, than if he instead would have met
a doctor who were not encouraged.

Unfortunately we do not have a randomized encouragement experiment. Instead
we are making use of observed differences in subscription pattern. For the
time being, let $Q$ be one if a patient is living in a \enquote*{high
prescription AA} county and $Q$ be zero if living in a \enquote*{low
prescription AA} county. Then (ii) holds by definition and as in the
randomized encouragement experiment, assumption (iii) is also likely to hold.

However it is not likely that assumption (i) holds. The reason is that high
and low prescription counties, may differ with respect to their
demography. This would imply that $Q\perp \! \! \! \perp (Y(1),Y(0))$ does not hold. As we have
substantial amounts of socioeconomic and health variables in our data, a
conditional version of assumption (i) could be valid.  Then, under the conditional independence%
\begin{equation}
Q\perp \! \! \! \perp (D(0),D(1),Y(1),Y(0))|\mathbf{X}  \label{CIA}
\end{equation}

\noindent we can use $Q$ as an instrument in the identification and
estimation of the effect of AA against ENZ given, (ii') $P(D=1|Q=1,\mathbf{X} )\neq P(D=1|Q=0,\mathbf{X})$ and (iii).

The so called IV estimators allow for confounding given that the confounding variables are
independent of $Q$ once we have controlled for $\mathbf{X}$.\footnote{Confounding variables are more precisely defined as variables not included in $\mathbf{X}$ that are simultaneously affecting treatment and outcome.} This possibility is enabled by the exclusion restriction given in assumption (\ref{CIA}).  Equation (\ref{CIA}) tells us that, given $\mathbf{X}$, $Q$ has no effect on the outcome $Y=Y(0)+D(Y(1)-Y(0))$ in the absence of individual treatment effects. Thus, $Q$ is excluded from the outcome regression under the null of no effect. In the case of  IV that is not randomized, the validity of this assumption can always be questioned as it cannot be formally tested. However, a  sensitivity analysis will be conducted evaluating assumption (\ref{CIA}).

Assumption (ii') tells us that $Q$ helps to predict prescription of AA and ENZ also given the pre-determined covariates. The relevance assumption (ii'), known as the first stage regression, is verified below.

Unfortunately we do not have institutional information enabling us to characterize low and high prescribing AA counties, what we observe is de facto subscription pattern of AA and ENZ in the 21 counties. Classifying the counties as being either low or high prescription AA
counties using the observed prescription is somewhat arbitrary, instead we use county as a factors as an instrument in the estimation. 

In this specific setting the identifying assumption (\ref{CIA}) mean that given the covariates there are no difference in health of the patients prescribed AA and ENZ across the county councils. Note that the design allow for unobserved differences in health across the counties but these differences in health cannot be correlated with the prescription pattern. As the quality of the health care affects the health and as it may be related to the prescription of the two drugs, we let historical county specific mortality related to prostate cancer be one of the covariates in the analysis. 

\subsection{Modeling the treatment assignment}

\label{sec:Validity}
Most often (or always) researchers using IV-estimators are estimating the first stage regression at the same time as conducting the analysis. The implication is that \enquote*{IV-estimators} could be susceptible to model specification errors. The reason is that the analysis, and hence design, many times becomes an iterative process as a consequence of observing irregularities in the data, finding estimates that are `strange' etc. The consequence is that p-values in testing for an effect will be flawed.  Here, the first stage and the test for the relevance of the county instruments take place before observing our outcomes and hence before the analysis takes place. The first stage sets restrictions on the outcome regression model which means that only one outcome regression for each outcome is possible. This makes our strategy comparable to a design based approach (cf. \cite{Imbens_Rubin_2015_causal}).\footnote{With design based estimators, e.g. matching, the design is specified before estimation takes place.}

The covariates included in $\mathbf{X}$ are the covariates displayed in Table \ref{sum_stat_intro} and nine factors constructed from a exploratory factor analysis of 126 continuous covariates. The results from the factor analysis are provided in Table \ref{FA} in Appendix. The first nine factor, which explain 43 percent of the variation of these covariates, will be added as covariates.\footnote{The first factor loads mainly on social security benefits of the patient and his family, the second on wages and the third on pensions and disposable income. The fourth factor loads on security benefits, the fifth mainly on sick leave, the sixth on private pensions. The seventh factor loads on the general health status and sickness progression of the individual. The eight factor loads mainly on pensions at diagnosis. The ninth factor loads on income from business.}

Let $c=1,...21$ index the 21 counties and let $q_{ic}=1$ if individual $i$ is living in county $c.$ Given the observed variation in the prescription across the 21 counties, it is reasonable to assume that the county where patient is living is also important for the type of prescribed drug in addition to the health of the patient as is described in $\mathbf{x}_{i}.$ We let the
latent propensity to receive AA be modeled as%
\begin{equation}
D_{i}^{\ast }=\mathbf{\gamma }_{x}^{\prime }\mathbf{x}_{i}+\gamma
_{c}q_{ic}+\varepsilon _{i}.  
\label{Latent D}
\end{equation}%
AA ($D_{i}=1)$ is prescribed if $D_{i}^{\ast }>0$ and ENZ ($D_{i}=0)$
if $D_{i}^{\ast }\leq 0.$ Let $\mathbf{q}_{i}=(q_{i1},...,q_{i21})$, $\mathbf{\gamma }_{C}=(\gamma _{1},...,\gamma
_{21})$ and $\mathbf{\gamma =(\gamma }_{x}^{\prime },\mathbf{%
\gamma }_{C})^{\prime },$ then $D_{i}^{\ast }=\mathbf{\gamma }%
^{\prime }\mathbf{z}_{i}+\varepsilon _{i}$, where $\mathbf{z}_{i}=(\mathbf{x}_{i}^{\prime},\mathbf{q}_{i}^{\prime})^{\prime}$. 

Under the assumption that $%
\varepsilon _{i}$ is normally distributed, the probability to be prescribed
AA can then be written 

\begin{equation}
\Pr (D_{i}=1|\mathbf{z}_{i},\mathbf{\mathbf{\gamma }})=\Pr (\varepsilon
_{i}>-\mathbf{\gamma }^{\prime }\mathbf{z}_{i})=\Phi (\mathbf{\gamma }%
^{\prime }\mathbf{z}_{i}),  
\label{PrD}
\end{equation}%
\noindent where $\Phi (.)$ is the cumulative distribution function of the
standard normal.

The results from the estimated probit regression model (\ref{PrD}) is presented in Table \ref{OLS}. From this table we can see that the county factor is the most important covariate in explaining the prescription of AA. We can however also see that patients with an acute myocardial infarction before the diagnosis have a lower probability of being prescribed AA. An overall test of relevance of the county factor is displayed in Table \ref{FTEST}. The F-statistic is $16.22$ with a p-value close to zero. In traditional IV analysis the rule of thumb for the relevance of IV is that the F-statistic should be above 10,  \citep{StaigerStock}. 

In order for the IV analysis to be valid we need to add the covariates displayed in Table \ref{OLS} in the three outcome regressions except for the county factor (i.e. the exclusion restriction on the outcome regression). This means the we have a model based design for the IV analysis and that given the model the p-values are correct, i.e. reflecting sampling uncertainty. 

\begin{table}[]
\caption{First stage probit regression}
\label{OLS}\centering 
{\footnotesize 
\begin{tabular}{@{\extracolsep{5pt}}lr} 
\textit{Dependent variable: Prescription of AA} \\
\hline \\[-1.8ex]
 ALDER\_T & 0.024 (0.049) \\ 
  I(ALDER\_T$\hat{\mkern6mu}$2) & $-$0.0001 (0.0003) \\ 
  Civil0 & $-$2.100 (1.947) \\ 
  Civil1 & $-$2.062 (1.948) \\ 
  UTBNFORGYMN & $-$0.112$^{*}$ (0.060) \\ 
  UTBNGYMN & $-$0.036 (0.055) \\ 
  Diff\_time & 0.00002 (0.00002) \\ 
  o\_tot1 & 0.040 (0.212) \\ 
  f\_tot1 & $-$0.105 (0.105) \\ 
  c\_tot1 & 0.481 (0.582) \\ 
  d\_tot1:ALDER\_T & $-$0.003 (0.002) \\ 
  d\_tot1:h\_tot1 & 0.098 (0.137) \\ 
  h\_tot1:ALDER\_T & 0.001 (0.007) \\ 
  c\_tot1:ALDER\_T & $-$0.005 (0.007) \\ 
  SCORE\_D23 & $-$0.010 (0.050) \\ 
  SCORE\_D24 & 0.075 (0.156) \\ 
  MEDS\_C07 & 0.013 (0.052) \\ 
  MEDS\_C08 & 0.041 (0.046) \\ 
  MEDS\_A10 & $-$0.072 (0.101) \\ 
  i211 & $-$0.239$^{***}$ (0.091) \\ 
  i481 & 0.106 (0.065) \\ 
  h\_tot\_ov1 & $-$0.150 (0.551) \\ 
  Factor1 & 0.016 (0.020) \\ 
  Factor2 & $-$0.003 (0.024) \\ 
  Factor3 & $-$0.042 (0.029) \\ 
  Factor4 & 0.032 (0.022) \\ 
  Factor5 & $-$0.023 (0.028) \\ 
  Factor6 & $-$0.075 (0.105) \\ 
  Factor7 & 0.034 (0.024) \\ 
  Factor8 & 0.015 (0.026) \\ 
  Factor9 & 0.013 (0.027) \\ 
  County:Dalarna & 0.528$^{***}$ (0.200) \\ 
  County:Gavleborg & $-$0.077 (0.217) \\ 
  County:Gotland & $-$0.154 (0.346) \\ 
  County:Halland & 0.118 (0.209) \\ 
  County:Jamtland & 0.516$^{**}$ (0.233) \\ 
  County:Jonkopings lan & 0.117 (0.207) \\ 
  County:Kalmar & 0.037 (0.223) \\ 
  County:Kronoberg & 1.273$^{***}$ (0.204) \\ 
  County:Norrbotten & 0.414$^{*}$ (0.213) \\ 
  County:Orebro & 0.356$^{*}$ (0.207) \\ 
  County:Ostergotlands lan & 0.071 (0.207) \\ 
  County:Skane & $-$0.425$^{**}$ (0.204) \\ 
  County:Sodermanland & $-$0.389 (0.250) \\ 
  County:Stockholm & 0.176 (0.200) \\ 
  County:Uppsala & 1.002$^{***}$ (0.197) \\ 
  County:Varmland & 0.055 (0.216) \\ 
  County:Vasterbotten & 0.516$^{**}$ (0.203) \\ 
  County:Vasternorrland & 0.651$^{***}$ (0.208) \\ 
  County:Vastmanland & 0.415$^{*}$ (0.214) \\ 
  County:Vastra gotalands lan & 0.561$^{***}$ (0.181) \\ 
    County\_specific\_mortality & $-$0.0005 (0.004) \\ 
   \hline
&  \\[-1.8ex] 
\multicolumn{2}{l}{Standard errors within parentheses. $^{*}p<$0.1; $^{**}p<$0.05; $^{***}p<$0.01.} \\ 
& 
\end{tabular}
}
\end{table}

\begin{table}[]
\caption{F-test from first stage probit regression, F-test of county factor}
\label{FTEST}\centering 
{\footnotesize 
\begin{tabular}{lllrrrr}
\hline \\[-.5ex] 
 \multicolumn{4}{l}{First stage probit regression} \\
  &&&& Df & F statistic & p-value \\ 
\hline \\[-3.0ex] 
  &  &  &  \\[-1.8ex]
 County factor ($\gamma_C$) &&&& 20& 16.24 & $<0.000$ \\ 
\hline\\[-0.5ex] 
\end{tabular}
}
\end{table}

\subsubsection{Sensitivity analyses for potential confounding using the first stage}
\label{sec:Sensitivty} 

The sensitivity analyses of the exclusion restriction of the county factor IV is conducted  by estimating the effects of the county factor IV on the two secondary outcomes, PAIN and SRE, in the period between diagnosis and prescription. If the county factor does  not contribute in explaining either of these two variables given that we control for $\mathbf{X}$, this will provide support for the exclusion restriction under the null of no differences in effects. 

PAIN is an indicator for severe pain which is defined as taken value one (zero else) if a patient simultaneously receive prescriptions of opiates in combination with Tramadol and Paracetamol (ATC-codes N02AA, N02AX02, and N02BE01) within a three month period between diagnosis and treatment. SRE is an indicator for a skeleton related event which is defined to be one if the patient experience a hospitalization because of a pathologic fracture (ICD codes M485, M495, M844 and M907) or spinal cord compression (ICD codes G550, G834, G952, G958, G959 and G992) \citep{Parry} between diagnosis and treatment. 

The importance of the county factor is examined using an F-test on $\mathbf{\alpha}_1=0$ based on maximum likelihood (ML) estimation of the probit regression model  

\begin{equation}
\Pr (Y_{i}=1|\mathbf{x}_{i},\mathbf{q}_{i})=\Phi (\mathbf{\alpha_0}^{\prime }\mathbf{x}_{i}+\mathbf{\alpha}_1^{\prime }\mathbf{q}_{i}),  
\label{PrY}
\end{equation}
where $Y_{i} $ is either PAIN or SRE.  As we are using two tests the overall risk of rejecting the validity of the the county factor IV will be higher than the nominal level in single tests of significance. A conservative strategy (as also will be used in the analysis of the three effects on the outcome) is to use the Bonferroni correction. Thus, with a five percent overall level, the levels of significance on the single test is 0.025.  

The results from this analysis is displayed in Table \ref{SRE}. The F-statistic for SRE and PAIN are 1.3526 and 1.0857, respectively. The corresponding p-values are 0.13 and 0.36. As these p-values are clearly above 0.025 we cannot reject the validity of the county factor. 

\begin{table}[]
\caption{F-test from probit regression on the covariates and the county factor, SRE and PAIN}
\label{SRE}\centering 
{\footnotesize 
\begin{tabular}{lllrrrr}
\hline \\[-.5ex] 
 \multicolumn{4}{l}{Sensitivity analysis I, prevalence of SRE} \\
  &&&& Df & F statistic & p-value \\ 
\hline \\[-3.0ex] 
  &  &  &  \\[-1.8ex]
 County ($\alpha_1$) &&&& 20& 1.3526  & 0.1345 \\ 
 \hline
\vspace{0.2in} \\
 \multicolumn{4}{l}{Sensitivity analysis II, prevalence of pain} \\
  &&&& Df & F statistic & p-value \\ 
\hline \\[-3ex] 
  &  &  &  \\[-1.8ex] 
 County ($\alpha_1$)& &&& 20 &   1.0857 & 0.3567 \\ 
\hline\\[-0.5ex] 
\multicolumn{7}{p{0.7\textwidth}}{Notes. Prevalence of SRE is an indicator that takes on value 1 if the individual have had one of the ICD codes specified between diagnose and treatment. According to this definition, 5 percent of the patients have had an SRE before treatment. Prevalence of pain is an indicator that takes on value 1 if the individual was prescribed opiates in combination with Paracetamol and Tramadol within 3 months, at one or more occasions between diagnose and treatment. According to this definition, about 3 percent of the patients have had severe pain before treatment. }
\end{tabular}
}
\end{table}

\subsection{The IV-design}

As we need to add covariates in our IV-analysis, model assumptions
are needed in the analysis. Furthermore as the outcomes (overall mortality, pain
and skeleton related events) are all indicator variables linear models are
not suitable in the analysis. For this reason we start by presenting linear
models for latent, or under-laying, health if prescribe AA or ENZ. After that, we explain how these latent health variables are related to the observed outcomes and, finally, we discuss how parameters in the model will  be estimated.

Let the unobserved health of individual $i$ if given ENZ or AA be 
\begin{equation*}
Y_{i}(0)^{\ast }=\beta _{0}+\mathbf{\delta }^{\prime }\mathbf{x}_{i}+u_{i0}
\end{equation*}%
and 
\begin{equation*}
Y_{i}(1)^{\ast }=\beta _{1}+\mathbf{\delta }^{\prime }\mathbf{x}_{i}+\mathbf{%
\delta }_{D}^{\prime }(\mathbf{x}_{i}-\overline{\mathbf{x}}\mathbf{)}+u_{i1},
\end{equation*}%
respectively. Here $\overline{\mathbf{x}}$ is the mean of the covariates in
the sample and $u_{i0}$ and $u_{i1}$ are error terms.
The unobserved health given the prescribed drug can then be formulated as a
function of the error terms and the observed covariates:%
\begin{equation}
Y_{i}^{\ast }=\beta _{0}+\mathbf{\delta }^{\prime }\mathbf{x}_{i}+\delta
_{1}D_{i}+\mathbf{\delta }_{D}^{\prime }D_{i}(\mathbf{x}_{i}-\overline{%
\mathbf{x}}\mathbf{)+}u_{i0}+D_{i}(u_{i1}-u_{i0}).  \label{Latent_reg}
\end{equation}%
With this specification $\delta _{1}=\beta _{1}-\beta _{0}$ is the average treatment effect on
the latent outcome and vector $\delta _{D}$ are heterogeneous effect,centered around $\delta _{1}$, with respect to the covariates. 

The potential problem estimating a treatment effect is that the doctors
could prescribe AA or ENZ based on health which we cannot observe $%
u_{i1}$ and $u_{i0}$. The selected prescription of the doctors imply in this
model that $\varepsilon _{i}$ in Equation (\ref{Latent D}) and $u_{i0}$ and $%
u_{i1}$ are correlated.

The estimation problem is solved by assuming $u_{i0}=\rho _{0}\varepsilon
_{i}+\eta _{i0}$ and $u_{i1}=\rho _{1}\varepsilon _{i}+\eta _{i1},$ where $%
\eta _{i0}$ and $\eta _{i1}$ are both random. This means that the unobserved
health is given as%
\begin{equation}
Y_{i}^{\ast }=\beta _{0}+\mathbf{\delta }^{\prime }\mathbf{x}_{i}+\delta
_{1}D_{i}+\mathbf{\delta }_{D}^{\prime }D_{i}(\mathbf{x}_{i}-\overline{%
\mathbf{x}}\mathbf{)+}\rho _{0}\varepsilon _{i}+\eta _{i0}+(\rho _{1}-\rho
_{0})D_{i}\varepsilon _{i}+D_{i}(\eta _{i1}-\eta _{i0}).
\label{latent}
\end{equation}%
Note that with this specification we assume that there is no unobserved heterogenity in the effects that are correlated with the county factor IV. This means that under model assumptions we identify the average treatment effect instead of the complier treatment effect.

In the data we observe $Y_{i}=1$ or $Y_{i}=0$ which then is used in the estimation of the parameters in  (\ref{latent}). To do this we need distribution assumptions on the error terms which provides the likelihood which will be maximized. The derivation of the likelihood is given in \citet[ch. 15]{Wooldridge_2002}, Chapter 15, but the main steps to form the likelihood is for transparency  given below.

We observe $Y_{i}=1$ if $Y_{i}^{\ast }\geq 0$ and $Y_{i}=0$ if $Y_{i}^{\ast
}<0$. Under the assumptions that $\varepsilon _{i},$ is independent of $%
\mathbf{z}_{i}$ and that $\varepsilon _{i}$, $\eta _{i1}$ and $\eta _{i0}$ are standard normal we get 
\begin{equation*}
\Pr (Y_{i}=1|\mathbf{z}_{i},D_{i}=1,\varepsilon _{i})=\Phi (\left[ \beta
_{0}+\mathbf{\delta }^{\prime }\mathbf{x}_{i}+\delta _{1}+\mathbf{\delta }%
_{D}^{\prime }(\mathbf{x}_{i}-\overline{\mathbf{x}}\mathbf{)+}\rho
_{1}\varepsilon _{i}\right] /(1-\rho _{1}^{2})^{1/2})
\end{equation*}

\noindent and 
\begin{equation*}
\Pr (Y_{i}=1|\mathbf{z}_{i},D_{i}=0,\varepsilon _{i})=\Phi (\left[ \beta
_{0}+\mathbf{\delta }^{\prime }\mathbf{x}_{i}+\rho _{0}\varepsilon _{i}%
\right] /(1-\rho _{0}^{2})^{1/2}.
\end{equation*}%
As 
\begin{equation*}
\Pr (\varepsilon _{i}|\varepsilon _{i}>-\mathbf{\gamma }^{\prime }\mathbf{z}%
_{i})=\phi (\varepsilon _{i})/\Phi (\mathbf{\gamma }^{\prime }\mathbf{z}%
_{i}),
\end{equation*}%
\noindent this means 
\begin{equation*}
\Pr (Y_{i}=1|\mathbf{z}_{i},D_{i}=1)=\frac{1}{\Phi (\mathbf{\gamma }^{\prime
}\mathbf{z}_{i})}\int_{-\mathbf{\gamma }^{\prime }\mathbf{z}_{i}}^{\infty
}\Phi (\left[ \beta _{0}+\mathbf{\delta }^{\prime }\mathbf{x}_{i}+\delta
_{1}+\mathbf{\delta }_{D}^{\prime }(\mathbf{x}_{i}-\overline{\mathbf{x}}%
\mathbf{)+}\rho _{1}\epsilon _{i}\right] /(1-\rho _{1})^{1/2})\phi (\epsilon
_{i})d\epsilon _{i}
\end{equation*}%
where $\epsilon $ in the integral is a dummy argument of integration.
Similarly 
\begin{equation*}
\Pr (Y_{i}=1|\mathbf{z}_{i},D_{i}=0)=\frac{1}{\Phi (\mathbf{-\gamma }%
^{\prime }\mathbf{z}_{i}))}\int_{-\infty }^{-\mathbf{\gamma }^{\prime }%
\mathbf{z}_{i}}\Phi (\left[ \beta _{0}+\mathbf{\delta }^{\prime }\mathbf{x}%
_{i}+\rho _{0}\epsilon _{i}\right] /(1-\rho _{0})^{1/2})\phi (\epsilon
_{i})d\epsilon _{i}
\end{equation*}

Define $p_i(\mathbf{\theta }_{1})=\Pr (Y_{i}=1|\mathbf{z}_{i},D_{i}=1)$ and $%
p_{i}(\mathbf{\theta }_{0})=\Pr (Y_{i}=1|\mathbf{z}_{i},D_{i}=0),$ where $%
\mathbf{\theta }_{1}\mathbf{=(}\beta _{0},\mathbf{\delta ,}\delta _{1},%
\mathbf{\delta }_{D},\mathbf{\gamma },\rho _{1})^{\prime }$ and $\mathbf{%
\theta }_{0}\mathbf{=(}\beta _{0},\mathbf{\delta ,\gamma },\rho
_{0})^{\prime },$ respectively. The likelihood to be maximized with respect
to $\mathbf{\theta =(}\beta _{0},\mathbf{\delta ,\delta _{1},\delta }_{D},%
\mathbf{\gamma },\rho _{0},\rho _{1})^{\prime }$ is then given as 
\begin{eqnarray*}
L(\mathbf{\theta };\mathbf{z}_{i},Y_{i},D_{i})=\prod_{i=1}^{n}\left[ p_{i}(%
\mathbf{\theta }_{1})^{Y_{i}}(1-p_{i}(\mathbf{\theta }_{1})^{(1-Y_{i})}\Phi (%
\mathbf{\gamma }^{\prime }\mathbf{z}_{i}\mathbf{)}\right] ^{D_{i}} \times \\ 
\left[p_{i}(\mathbf{\theta }_{0})^{Y_{i}}(1-p_{i}(\mathbf{\theta }%
_{0})^{(1-Y_{i})}(1-\Phi (\mathbf{\gamma }^{\prime }\mathbf{z}_{i}\mathbf{)}%
\right] ^{(1-D_{i})}. 
\label{Likelihood}
\end{eqnarray*}

The conditional individual treatment effect, that is the effect conditional on $\mathbf{x}_{i}$, can be estimated by using the differences in the expected outcome under treatment minus the expected outcome if not treated. Thus,  %
\begin{equation}
\widehat{\Delta }(\mathbf{x}_{i})=\Phi (\widehat{\beta }_{0}+\widehat{%
\mathbf{\delta }}^{\prime }\mathbf{x}_{i}+\widehat{\delta }_{1}+\widehat{%
\mathbf{\delta }}_{D}^{\prime }(\mathbf{x}_{i}-\overline{\mathbf{x}}\mathbf{%
))-}\Phi (\widehat{\beta }_{0}+\widehat{\mathbf{\delta }}^{\prime }\mathbf{x}%
_{i}\mathbf{),}
\label{CTE_IV}
\end{equation}%
where the $^{\widehat{}}$ is the maximum likelihood estimates. The conditional average
treatment effect (CATE) is then estimated as 
\begin{equation}
\widehat{\Delta }=\sum_{i=1}^{n}\widehat{\Delta }(\mathbf{x}_{i})
\label{ATE_IV}
\end{equation}%
\bigskip 

\section{Planned analysis}

\label{sec:Plan}

The same analyses as in \cite{Johansson_et_al_2021a} will be conducted. \ This
means that we have one primary, and two secondary outcomes. The primary
outcome is all-cause mortality (DEAD) and the two secondary outcomes
capturing morbidity are PAIN, and SRE defined in Section \ref{sec:Sensitivty}.

Bonferroni correction with a five percent overall level will be used in the inferences. With one primary and two
secondary outcomes this means that the significance level on the single outcomes will be 1.67\% $(=100\ast 0.05/3)$.

DEAD as a primary endpoint is an indicator variable defined as one for
patients who are dead of any causes and zero for other patients at the end of each 30 days period after being prescribed AA or ENZ. Mortality data will be available until the end of June, 2020. Consequently, the first
patient administered the treatments can theoretically have up to 70
mortality registrations, but the actual number of registrations will most
likely be fewer as they are suffer a deadly disease.

Prescription data measuring PAIN will be available until 31st of December, 2020. Therefore, up to 76 periods are available for estimating the effects on PAIN. As inpatient care data is available until 31st of December, 2019, at most 64 periods are available measuring effects on SRE.

For each of the periods $t=1,...T$, we will estimate conditional individual treatment defined in (\ref{CTE_IV}) and then we estimate the average treatment effect given in (\ref{ATE_IV}).  To be specific, we first estimate 
\begin{equation*}
\widehat{\Delta }(\mathbf{x}_{i},t)=\Phi (\widehat{\beta }_{0t}+\widehat{%
\mathbf{\delta }}_{t}^{\prime }\mathbf{x}_{i}+\widehat{\delta }_{1t}+%
\widehat{\mathbf{\delta }}_{Dt}^{\prime }(\mathbf{x}_{i}-\overline{\mathbf{x}%
}\mathbf{))-}\Phi (\widehat{\beta }_{0t}+\widehat{\mathbf{\delta }}%
_{t}^{\prime }\mathbf{x}_{i}\mathbf{),}
\end{equation*}%
where $\widehat{\beta }_{0t},\widehat{\mathbf{\delta }}_{t}^{\prime }\mathbf{%
,}\widehat{\delta }_{1t}$ and $\widehat{\mathbf{\delta }}_{Dt}^{\prime }$,
are the maximum likelihood estimates from the model at each time period $t$.
The average treatment effect at each month $t$ is estimated as 
\begin{equation}
\widehat{\Delta }_{t}=\sum_{i=1}^{n}\widehat{\Delta }(\mathbf{x}_{i},t)
\end{equation}

We will display Bonferroni adjusted confidence intervals (i.e. using the 1.67\% level for an overall 5\% level test) at each time period. Standard errors of $\widehat{\Delta }_{t}$ will be estimated using bootstrapping.

The problem with analysis of the two morbidity outcomes is that they are only observed in the data if the patient is alive. In the analysis above, at each evaluation period, we will exclude the dead patients from the analysis. This means that the number of valid observations will be reduced over the evaluation window. If there are no differences in mortality rates between the two drugs, this analysis strategy provides unbiased estimates of the comparative effectiveness on morbidity. However, if there are differences in mortality rates, these analyses are biased, as it is more likely that we will observe a morbidity for the drug with lower mortality rates. If this is the case, we will need to estimate bounds of potential effects as a sensitivity analysis. We labeled all patients who died as either having the morbidity or not (i.e. PAIN =1 and SRE=1 and  PAIN =0 and SRE=0). If the mortality is observed to be higher for ENZ, the first case provides the upper bound estimate of the comparative effectiveness of AA against ENZ while the second one provides the lower bound on these two morbidity outcomes, and vice versa if the mortality rate is lower for AA.

\subsection{Sensitivity analysis}
All observational studies can give biased results. In this case, the problem is that the model used in the analysis is incorrectly specified. For this reason placebo regression will be estimated, that is we will estimate effect where we should not find effects given the design is valid. 

We will add data containing detailed information on patient's heath with regard to the prostate cancer from the national prostate cancer register (NPCR). The data is ordered and we will have access to it at the same time
as we have the mortality data. 

For the test we will use three covariates, that during discussions with specialists, are judged to be important confounders: PSA levels, Gleason score, and metastases at the time of prostate cancer diagnoses. These three pre-treatment covariates will be used as outcomes in the same regression as for the main analysis. If we find a statistical significant effect on these outcome this is signal of hidden bias in our analyses. 

With three pre-measured covariates we, as in the main analysis, adjust the significance level for the individual tests using Bonferroni correction based on a five percent overall level. This means that
the level for testing on each single outcome is 1.67 percent.

\subsection{A potential survival analysis}
We choose to model the prevalence of mortality as the outcomes mainly because of previous experience using this model in IV analysis. However, we will alsoexplore the possibilities of estimating an overall effect on mortality using duration analysis. This section presents the idea of the IV analysis on a duration outcome. The reason for not explicitly add it as test of an effect is that we have not found any similar IV analysis in the literature. Our experience will be discussed in the the paper presenting the results. 

Let $Y_{it}=1$ if patient $i$ dies in time period $t$. Furthermore, let $p_{i}(\mathbf{\theta }_{1t})=\Pr (Y_{it}=1|\mathbf{z}_{i},D_{i}=1,Y_{it-1}=0)$ be the probability that patient $i$ prescribed AA dies at month $t$ given survival up to this month and let $p_{i}(\mathbf{\theta }_{0t})=\Pr (Y_{i}=1|\mathbf{z}_{i},D_{i}=0,Y_{it-1}=0)$ be the corresponding conditional probability if the patient instead was prescribed ENZ. Assume 

\begin{equation*}
p_{i}(\mathbf{\theta }_{1t})=\frac{1}{\Phi (\mathbf{\gamma }^{\prime
}\mathbf{z}_{i})}\int_{-\mathbf{\gamma }^{\prime }\mathbf{z}_{i}}^{\infty
}\Phi (\left[ \beta _{t}+\mathbf{\delta }^{\prime }\mathbf{x}_{i}+\delta
_{1}+\mathbf{\delta }_{D}^{\prime }(\mathbf{x}_{i}-\overline{\mathbf{x}}%
\mathbf{)+}\rho _{1}\epsilon _{i}\right] /(1-\rho _{1})^{1/2})\phi (\epsilon
_{i})d\epsilon _{i}
\end{equation*}%
and 

\begin{equation*}
p_{i}(\mathbf{\theta }_{0t})=\frac{1}{(1-\Phi (\mathbf{\gamma }%
^{\prime }\mathbf{z}_{i}))}\int_{-\infty }^{-\mathbf{\gamma }^{\prime }%
\mathbf{z}_{i}}\Phi (\left[ \beta _{t}+\mathbf{\delta }^{\prime }\mathbf{x}%
_{i}+\rho _{0}\epsilon _{i}\right] /(1-\rho _{0})^{1/2})\phi (\epsilon
_{i})d\epsilon _{i}.
\end{equation*}%
Thus  $\mathbf{\theta }_{1t}\mathbf{=(}\beta _{t},\mathbf{\delta ,}\delta _{1},\mathbf{\delta }_{D},\mathbf{\gamma },\rho _{1})^{\prime }$ and $\mathbf{\theta }_{0t}\mathbf{=(}\beta _{t},\mathbf{\delta ,\gamma },\rho_{0})^{\prime }$. $\beta _{t} + \mathbf{\gamma }^{\prime }\mathbf{z}_{i}$ measures the base line conditional probability to die given survival up to month $t$. $\delta_1$ and $\mathbf{\delta }_{D}^{\prime }(\mathbf{x}_{i}-\overline{\mathbf{x}})$ measure the `shift' in this baseline probability, i.e., an effect. Thus we restrict the effect on the conditional probability to be the same at all months.

Let  $\Xi =(\beta_1,....,\beta_{T-1})^{\prime}$, where $T$ is the last follow up month and let $T_i$ be the number of month the individuals is alive in the the data, thus $T_i=T$, if the patient is alive when we end the study. The likelihood to be maximized with respect to $\mathbf{\theta} =(\Xi,\delta ,\delta _{1},\delta_{D},\mathbf{\gamma },\rho _{0},\rho _{1})^{\prime }$ is
\begin{eqnarray*}
L(\mathbf{\theta };\mathbf{z}_{i},Y_{it},D_{i})=\prod_{i=1}^{n}\prod_{t=1}^{T_i}\left[ p_{i}(%
\mathbf{\theta }_{1t})^{Y_{it}}(1-p_{i}(\mathbf{\theta }_{1t})^{(1-Y_{it})}\Phi (%
\mathbf{\gamma }^{\prime }\mathbf{z}_{i}\mathbf{)}\right] ^{D_{i}} \times \\
\left[p_{i}(\mathbf{\theta }_{0t})^{Y_{it}}(1-p_{i}(\mathbf{\theta }%
_{0t})^{(1-Y_{it})}(1-\Phi (\mathbf{\gamma }^{\prime }\mathbf{z}_{i}\mathbf{)}%
\right] ^{(1-D_{i})}.
\end{eqnarray*}

Based on the maximum likelihood estimates (i.e. $\widehat{\mathbf{\theta }}$) we can estimate the survival function up to any month $t$ for the AA and ENZ patients, respectively. Let $R_{1t}$ and $R_{0t}$ be the risk set of the AA and ENZ patients in period $t$, respectively. These risk sets contains $n_{1t}$ and $n_{0t}$ patients that has not yet died. The estimated survival function up to $t$ are then 

\begin{eqnarray*}
S(\widehat{\mathbf{\theta }},D=1,t) = \frac{1}{n_{1t}}\sum_{i:R_{1t}}\prod_{\tau=1}^{t}(1-p_{i}(\widehat{\mathbf{\theta }}_{1\tau}))
\end{eqnarray*}
and
\begin{eqnarray*}
S(\widehat{\mathbf{\theta }},D=0,t) = \frac{1}{n_{0t}}\sum_{i:R_{0t}}\prod_{\tau=1}^{t}(1-p_{i}(\widehat{\mathbf{\theta }}_{0\tau})),
\end{eqnarray*}
respectively.

The overall effect on survival up to a given period $\bar{T}$ is then estimated as

\begin{equation}
\widehat{\Delta }_{\bar{T}}=\sum_{t=1}^{\bar{T}}S(\widehat{\mathbf{\theta }},D=1,t)-S(\widehat{\mathbf{\theta }},D=0,t).
\end{equation}

\section{Discussion}

\label{sec:Discussion}

The paper present a study protocol, or instrumental variables (IV) design, for an IV-analysis of the
comparative effect of abiraterone acetate (AA) against enzalutamide (ENZ) for prostate cancer patients on overall mortality, pain and skeleton related events. The design can be seen as a `natural experiment', that is we have an IV that: (1) alters the probability of treatment at the individual level, but is (2) not otherwise correlated with the outcome. In this setting we make use of differences in prescription practices across 21 Swedish county councils as a `county factor' IV. 

Assumption (2) may not be valid as it is possible that health and quality in health care differs across the 21 county councils in Sweden. For this reason detailed data on observed health and socioeconomic status of the patients, taken from from population registers, is included in the design. In addition, mortality data in prostate cancer at the county level is included in the design. The relevance of the county factor IV given covariates (i.e. requirement (1) given $\mathbf{X}$) is confirmed in data, that is, the county factor IV provides a sufficient shift in the individual probability to be prescribed AA or ENZ, given the observed covariates.

Conditional (on $\mathbf{X}$) IV-analysis could be flawed as the analysis to some extent needs parametric model assumptions. If the researcher observe the outcome there is a risk that the researcher conflate the design with analysis. That is, there is a risk of choosing the model specification (by e.g. adding or removing covariates) to obtain statistical significant result. By making the IV-design public before observing the outcomes this modeling bias is controlled for. According to our understanding, this is the first study that publish an IV design in a pre-analysis plan and this even before observing our main outcome, mortality.  

Another problem  with conditional IV-analysis is that we do not know which covariates to add to make the IV valid. For this reason we conduct a sensitivity analysis of the validity of the IV-design (i.e. assumption (2) given $\mathbf{X}$) by using data on the two morbidity outcomes (pain and skeleton related events) observed before prescription date. We estimate the comparative effect of the county factor on these outcomes on data in the period between diagnosis and prescription. As the county factor did not contribute in explaining these two pre-measured outcomes the invalidity of the county IV could \textit{not} be rejected.   

The design makes use of data from \cite{Johansson_et_al_2021a} who used a Rubin Causal Model design (see \cite{Imbens_Rubin_2015_causal}) in their design for the same comparative effectiveness evaluation. That is, the design assumes that observed covariates on health and socioeconomic status controls for all confounding. The IV-design allows for confounding but relays on the other hand on the exclusion restriction assumption (2). As both design could be valid comparison of the results from the two studies will be of interest and will also provide an understanding of pros and cons of the two type of designs.

\newpage

\bibliographystyle{agsm}
\bibliography{Refs2}

\newpage

\section*{Appendix}
\label{sec:Appendix}

\begin{table}[!htbp] \centering
\ContinuedFloat
  \caption{Exploratory factor analysis} 
    \label{FA} 
{\footnotesize 
\centerline{
\begin{tabular}{rrrrrrrrrr}
  \hline
 & Factor1 & Factor2 & Factor3 & Factor4 & Factor5 & Factor6 & Factor7 & Factor8 & Factor9 \\ 
  \hline
SocBidrPers\_1y\_D & 0.94 &  &  &  &  &  &  &  &  \\ 
  SocBidrPers\_D & 0.93 &  &  &  &  &  &  &  &  \\ 
  SocBidrFam\_D & 0.92 &  &  &  &  &  &  &  &  \\ 
  SocBidrPers\_2y\_D & 0.91 &  &  &  &  &  &  &  &  \\ 
  SocBidrFam\_2y\_D & 0.90 &  &  &  &  &  &  &  &  \\ 
  SocBidrFam\_1y\_D & 0.90 &  &  &  &  &  &  &  &  \\ 
  SocBidrPersF04\_2y & 0.88 &  &  &  &  &  &  &  &  \\ 
  SocBidrFam\_2y & 0.87 &  &  &  &  &  &  &  &  \\ 
  SocBidrPersF04\_1y & 0.87 &  &  &  &  &  &  &  &  \\ 
  SocBidrFam & 0.86 &  &  &  &  &  &  &  &  \\ 
  SocBidrPersF04 & 0.85 &  &  &  &  &  &  &  &  \\ 
  SocBidrFam\_1y & 0.84 &  &  &  &  &  &  &  &  \\ 
  LoneInk\_1y &  & 0.95 & 0.22 &  &  &  &  &  &  \\ 
  LoneInk.x &  & 0.94 &  &  &  &  &  &  &  \\ 
  ForvErs\_1y &  & 0.94 & 0.23 &  &  &  &  &  &  \\ 
  ForvErs &  & 0.94 & 0.20 &  &  &  &  &  &  \\ 
  LoneInk\_2y &  & 0.92 & 0.27 &  &  &  &  &  &  \\ 
  ForvErs\_2y &  & 0.90 & 0.28 &  &  &  &  &  &  \\ 
  LoneInk\_1y\_D &  & 0.60 & 0.58 &  &  &  &  & -0.38 &  \\ 
  LoneInk\_D &  & 0.59 & 0.56 &  &  &  &  & -0.37 &  \\ 
  LoneInk\_2y\_D &  & 0.44 & 0.62 &  &  &  &  & -0.39 &  \\ 
  DispInk04 &  & 0.31 & 0.37 &  &  &  &  &  &  \\ 
  DispInk04\_1y &  & 0.25 & 0.34 &  &  &  &  &  &  \\ 
  DispInk\_D &  & 0.24 & 0.53 &  &  &  &  &  &  \\ 
  DispInkFam04\_1y &  & 0.21 & 0.32 &  &  &  &  &  &  \\ 
  SumTjP\_2y &  & -0.21 & 0.85 &  &  &  &  & 0.30 &  \\ 
  SumAldP03 &  & -0.29 &  &  & -0.51 &  &  & 0.27 &  \\ 
  SumAldP03\_2y &  & -0.30 &  &  & -0.53 &  &  & 0.29 &  \\ 
  SumAldP03\_1y &  & -0.31 &  &  & -0.54 &  &  & 0.28 &  \\ 
  SumTjP\_1y &  &  & 0.88 &  &  &  &  & 0.27 &  \\ 
  SumTjP &  &  & 0.85 &  &  &  &  & 0.26 &  \\ 
  SumTjp\_D &  &  & 0.60 &  &  &  &  & 0.68 &  \\ 
  SumTjp\_1y\_D &  &  & 0.51 &  &  &  &  & 0.77 &  \\ 
  SumTjp\_2y\_D &  &  & 0.41 &  &  &  &  & 0.77 &  \\ 
  AldPens\_D &  &  & 0.35 &  & -0.21 & 0.37 &  & 0.75 &  \\ 
  DispInk\_2y\_D &  &  & 0.32 &  &  &  &  &  &  \\ 
  DispInk\_1y\_D &  &  & 0.30 &  &  &  &  &  &  \\ 
  DispInkFam\_D &  &  & 0.30 &  &  &  &  &  &  \\ 
  PrivPens &  &  & 0.29 &  &  & 0.58 &  &  &  \\ 
  DispInkFam\_2y\_D &  &  & 0.26 &  &  &  &  &  &  \\ 
  AldPens\_1y\_D &  &  & 0.26 &  &  & 0.37 &  & 0.81 &  \\ 
  DispInkFam04\_2y &  &  & 0.25 &  &  &  &  &  &  \\ 
  DispInkFam\_1y\_D &  &  & 0.25 &  &  &  &  &  &  \\ 
  DispInk04\_2y &  &  & 0.25 &  &  &  &  &  &  \\  
    PrivPens\_2y &  &  & 0.24 &  &  & 0.92 &  &  &  \\ 
  PrivPens\_1y &  &  & 0.24 &  &  & 0.91 &  &  &  \\ 
  DispInkFam04 &  &  & 0.23 &  &  &  &  &  &  \\ 
     \hline
\end{tabular}
}}
\end{table}

\begin{table}[!htbp] \centering
\ContinuedFloat
  \caption{Exploratory factor analysis} 
      \label{FA}
      {\footnotesize 
\centerline{
\begin{tabular}{rrrrrrrrrr}
  \hline
 & Factor1 & Factor2 & Factor3 & Factor4 & Factor5 & Factor6 & Factor7 & Factor8 & Factor9 \\ 
  \hline
  ForTid\_1y\_D &  &  &  & 0.95 &  &  &  &  &  \\ 
  ForTid\_D &  &  &  & 0.91 &  &  &  &  &  \\ 
  ForTid\_2y\_D &  &  &  & 0.85 &  &  &  &  &  \\ 
  SocInk\_1y\_D &  &  &  & 0.83 &  &  &  &  &  \\ 
  SocInk\_2y\_D &  &  &  & 0.79 &  &  &  &  &  \\ 
  SocInk\_D &  &  &  & 0.67 & 0.31 &  &  &  &  \\ 
  ForTid\_2y &  &  &  & 0.59 & 0.35 &  &  &  &  \\ 
  ForTid\_1y &  &  &  & 0.55 & 0.43 &  &  &  &  \\ 
  ForTid &  &  &  & 0.50 & 0.45 &  &  &  &  \\ 
  SocInk\_2y &  &  &  & 0.43 & 0.67 &  &  &  &  \\ 
  SocInk\_1y &  &  &  & 0.36 & 0.85 &  &  &  &  \\ 
  SocInk &  &  &  & 0.23 & 0.82 &  &  &  &  \\ 
  SjukRe\_2y\_D &  &  &  & 0.23 &  &  &  &  &  \\ 
  SjukRe\_1y &  &  &  &  & 0.72 &  &  &  &  \\ 
  SjukRe &  &  &  &  & 0.68 &  &  &  &  \\ 
  SjukRe\_2y &  &  &  &  & 0.58 &  &  &  &  \\ 
  SjukRe\_D &  &  &  &  & 0.45 &  &  &  &  \\ 
  nn &  &  &  &  & 0.24 &  & 0.42 &  &  \\ 
  PrivPens\_1y\_D &  &  &  &  &  & 0.94 &  &  &  \\ 
  PrivPens\_D &  &  &  &  &  & 0.93 &  &  &  \\ 
  PrivPens\_2y\_D &  &  &  &  &  & 0.93 &  &  &  \\ 
  AldPens\_2y\_D &  &  &  &  &  & 0.37 &  & 0.81 &  \\ 
  bft\_12m\_tot &  &  &  &  &  &  & 0.93 &  &  \\ 
  bft\_12m\_tot\_c619 &  &  &  &  &  &  & 0.61 &  &  \\ 
  bft\_3m &  &  &  &  &  &  & 0.61 &  &  \\ 
  bft\_4m &  &  &  &  &  &  & 0.59 &  &  \\ 
  bft\_5m &  &  &  &  &  &  & 0.58 &  &  \\ 
  bft\_2m &  &  &  &  &  &  & 0.57 &  &  \\ 
  q4\_t &  &  &  &  &  &  & 0.54 &  &  \\ 
  bft\_1m &  &  &  &  &  &  & 0.52 &  &  \\ 
  q3\_t &  &  &  &  &  &  & 0.44 &  &  \\ 
  c\_tot\_bw &  &  &  &  &  &  & 0.35 &  &  \\ 
  q2\_t &  &  &  &  &  &  & 0.34 &  &  \\ 
  msv\_days\_c619 &  &  &  &  &  &  & 0.34 &  &  \\ 
  msv\_tot\_prop &  &  &  &  &  &  & 0.33 &  &  \\ 
  bfd\_12m &  &  &  &  &  &  & 0.33 &  &  \\ 
  bfd\_60m &  &  &  &  &  &  & 0.33 &  &  \\ 
  msv\_days &  &  &  &  &  &  & 0.32 &  &  \\ 
  msv\_tot\_prop\_c619 &  &  &  &  &  &  & 0.29 &  &  \\ 
  h\_bft\_3m &  &  &  &  &  &  & 0.25 &  &  \\ 
  q1\_t &  &  &  &  &  &  & 0.24 &  &  \\ 
  h\_tot\_bw &  &  &  &  &  &  & 0.22 &  &  \\ 
  InkFNetto\_1y &  &  &  &  &  &  &  &  & 0.97 \\ 
  InkFNetto\_2y &  &  &  &  &  &  &  &  & 0.96 \\ 
  InkFNetto &  &  &  &  &  &  &  &  & 0.95 \\ 
  InkFNetto\_1y\_D &  &  &  &  &  &  &  &  & 0.73 \\ 
  InkFNetto\_2y\_D &  &  &  &  &  &  &  &  & 0.71 \\ 
  InkFNetto\_D &  &  &  &  &  &  &  &  & 0.70 \\  
     \hline
SS loadings & 9.56 & 6.96 & 6.50 & 5.72 & 5.48 & 5.30 & 4.86 & 4.80 & 4.39 \\ 
  Proportion Var & 0.08 & 0.06 & 0.05 & 0.05 & 0.04 & 0.04 & 0.04 & 0.04 & 0.03 \\ 
  Cumulative Var & 0.08 & 0.13 & 0.18 & 0.23 & 0.27 & 0.31 & 0.35 & 0.39 & 0.43 \\ 
\end{tabular}
}}
\end{table}

\begin{table}[ht]
\centering
\footnotesize
\caption{ICD codes used to specify related diagnoses and medicines}
\label{diseases}
\begin{tabular}{llllll}
\\[-1.8ex]\hline 
\\[-1.8ex]
& Cardiovascular & Diabetes & Osteoporosis & Metastases & Malaise and \\
&disease&&&  & fatigue \\ 
\hline 
\\[-1.8ex]
Diagnoses & I21 & E10 & M859 & C77 & R53 \\ 
& I22 & E11 & M810 & C78 \\ 
& I252 & E12 & M818 & C79\\ 
& Z958 & E13 & M819 \\ 
& Z959 & E14 & \\ 
& I70 & & \\ 
& I71 & & \\ 
& I731 & & \\ 
& I738 & & \\ 
& I739 & & \\ 
& I771 & & \\ 
& I790 & & \\ 
& I792 & & \\ 
& K551 & & \\ 
& K558 & & \\ 
& K559 & & \\ 
& I090 & & \\ 
& I110 & & \\ 
& I13 & & \\ 
& I130 & & \\ 
& I131 & & \\ 
& I132 & & \\ 
& I139 & & \\ 
& I42 & & \\ 
& I420 & & \\ 
& I421 & & \\ 
& I422 & & \\ 
& I423 & & \\ 
& I424 & & \\ 
& I425 & & \\ 
& I426 & & \\ 
& I427 & & \\ 
& I428 & & \\ 
& I429 & & \\ 
& I43 & & \\ 
& I44 & & \\ 
& I45 & & \\ 
& I46 & & \\ 
& I47 & & \\ 
& I48 & & \\ 
& I49 & & \\ 
& I50 & & \\ 
& I51 & & \\ 
& R00 & & \\ 
\hline
\\[-1.8ex]
Medicines & C07 &A10 & \\ 
& C08 & & \\ 
\hline 
\end{tabular}
\end{table}

\begin{table}[!htbp] \centering
\footnotesize 
  \caption{Sensitivity analysis I} 
  \label{Compl_samp2} 
\begin{tabular}{@{\extracolsep{5pt}}lr} 
\textit{Dependent variable: Prevalence of SRE} \\
\hline \\[-1.8ex]
 ALDER\_T & 0.054 (0.191) \\ 
  I(ALDER\_T$\hat{\mkern6mu}$2) & $-$0.0003 (0.001) \\ 
  Civil0 & $-$4.085 (6.729) \\ 
  Civil1 & $-$4.117 (6.710) \\ 
  UTBNFORGYMN & 0.006 (0.123) \\ 
  UTBNGYMN & $-$0.011 (0.094) \\ 
  Diff\_time & 0.00001 (0.00003) \\ 
  o\_tot1 & 1.020$^{***}$ (0.179) \\ 
  f\_tot1 & 0.092 (0.121) \\ 
  c\_tot1 & 1.379 (2.243) \\ 
  d\_tot1:ALDER\_T & $-$0.004 (0.003) \\ 
  d\_tot1:h\_tot1 & 0.105 (0.209) \\ 
  h\_tot1:ALDER\_T & 0.015 (0.012) \\ 
  c\_tot1:ALDER\_T & $-$0.010 (0.028) \\ 
  SCORE\_D23 & 0.054 (0.080) \\ 
  SCORE\_D24 & $-$0.489 (0.407) \\ 
  MEDS\_C07 & 0.023 (0.085) \\ 
  MEDS\_C08 & 0.003 (0.075) \\ 
  MEDS\_A10 & 0.216 (0.161) \\ 
  i211 & $-$0.031 (0.134) \\ 
  i481 & $-$0.025 (0.108) \\ 
  h\_tot\_ov1 & $-$1.137 (0.910) \\ 
  Factor1 & $-$0.934 (20.517) \\ 
  Factor2 & $-$0.007 (0.047) \\ 
  Factor3 & 0.035 (0.514) \\ 
  Factor4 & 0.007 (0.041) \\ 
  Factor5 & 0.050 (0.599) \\ 
  Factor6 & $-$0.022 (0.147) \\ 
  Factor7 & 0.204$^{***}$ (0.035) \\ 
  Factor8 & 0.035 (0.319) \\ 
  Factor9 & $-$0.011 (0.041) \\ 
  lan\_namnDalarna & $-$0.006 (0.588) \\ 
  lan\_namnGavleborg & 0.204 (0.534) \\ 
  lan\_namnGotland & 0.075 (0.621) \\ 
  lan\_namnHalland & 0.329 (0.524) \\ 
  lan\_namnJamtland & 0.490 (0.711) \\ 
  lan\_namnJonkopings lan & 0.241 (0.535) \\ 
  lan\_namnKalmar & 0.229 (0.531) \\ 
  lan\_namnKronoberg & 0.420 (0.533) \\ 
  lan\_namnNorrbotten & 0.525 (0.535) \\ 
  lan\_namnOrebro & $-$0.259 (0.709) \\ 
  lan\_namnOstergotlands lan & 0.094 (0.537) \\ 
  lan\_namnSkane & 0.081 (0.510) \\ 
  lan\_namnSodermanland & 0.180 (0.530) \\ 
  lan\_namnStockholm & $-$0.025 (0.514) \\ 
  lan\_namnUppsala & $-$0.095 (0.542) \\ 
  lan\_namnVarmland & 0.522 (0.517) \\ 
  lan\_namnVasterbotten & 0.260 (0.531) \\ 
  lan\_namnVasternorrland & 0.758 (0.511) \\ 
  lan\_namnVastmanland & $-$0.066 (0.594) \\ 
  lan\_namnVastra gotalands lan & 0.351 (0.500) \\ 
Country\_specific\_mortality & $-$0.015$^{**}$ (0.008) \\
    \hline \\[-1.8ex] 
 \multicolumn{2}{l}{$^{*}$p$<$0.1; $^{**}$p$<$0.05; $^{***}$p$<$0.01} \\ 
    \end{tabular} 
\end{table}

\begin{table}[!htbp] \centering
\footnotesize 
  \caption{Sensitivity analysis II} 
  \label{Compl_samp2} 
\begin{tabular}{@{\extracolsep{5pt}}lr} 
\textit{Dependent variable: Prevalence of pain} \\
\hline \\[-1.8ex]
 ALDER\_T & $-$0.004 (0.082) \\ 
  I(ALDER\_T$\hat{\mkern6mu}$2) & $-$0.00003 (0.001) \\ 
  Civil0 & $-$1.918 (3.280) \\ 
  Civil1 & $-$1.907 (3.282) \\ 
  UTBNFORGYMN & 0.231$^{**}$ (0.114) \\ 
  UTBNGYMN & 0.120 (0.113) \\ 
  Diff\_time & 0.0001$^{*}$ (0.00003) \\ 
  o\_tot1 & 0.062 (0.329) \\ 
  f\_tot1 & $-$0.248 (0.221) \\ 
  c\_tot1 & 0.491 (1.182) \\ 
  d\_tot1:ALDER\_T & 0.002 (0.002) \\ 
  c\_tot1:h\_tot1 & 0.062 (0.184) \\ 
  h\_tot1:ALDER\_T & $-$0.004 (0.012) \\ 
  c\_tot1:ALDER\_T & $-$0.004 (0.015) \\ 
  SCORE\_D23 & 0.127 (0.087) \\ 
  SCORE\_D24 & 0.077 (0.250) \\ 
  MEDS\_C07 & 0.021 (0.093) \\ 
  MEDS\_C08 & $-$0.072 (0.084) \\ 
  MEDS\_A10 & $-$0.012 (0.162) \\ 
  i211 & $-$0.009 (0.132) \\ 
  i481 & 0.109 (0.104) \\ 
  h\_tot\_ov1 & 0.347 (0.871) \\ 
  Factor1 & 0.071$^{***}$ (0.013) \\ 
  Factor2 & $-$0.025 (0.097) \\ 
  Factor3 & 0.010 (0.047) \\ 
  Factor4 & 0.031 (0.030) \\ 
  Factor5 & $-$0.020 (0.044) \\ 
  Factor6 & 0.008 (0.021) \\ 
  Factor7 & 0.111$^{***}$ (0.022) \\ 
  Factor8 & 0.061 (0.044) \\ 
  Factor9 & $-$0.064$^{***}$ (0.015) \\ 
  lan\_namnDalarna & 0.005 (0.361) \\ 
  lan\_namnGavleborg & 0.134 (0.342) \\ 
  lan\_namnGotland & 0.303 (0.423) \\ 
  lan\_namnHalland & $-$0.258 (0.430) \\ 
  lan\_namnJamtland & 0.352 (0.374) \\ 
  lan\_namnJonkopings lan & $-$0.144 (0.388) \\ 
  lan\_namnKalmar & $-$0.009 (0.385) \\ 
  lan\_namnKronoberg & $-$0.039 (0.404) \\ 
  lan\_namnNorrbotten & 0.148 (0.371) \\ 
  lan\_namnOrebro & 0.001 (0.384) \\ 
  lan\_namnOstergotlands lan & $-$0.118 (0.380) \\ 
  lan\_namnSkane & 0.185 (0.321) \\ 
  lan\_namnSodermanland & 0.001 (0.369) \\ 
  lan\_namnStockholm & $-$0.296 (0.361) \\ 
  lan\_namnUppsala & $-$0.304 (0.431) \\ 
  lan\_namnVarmland & 0.114 (0.353) \\ 
  lan\_namnVasterbotten & 0.373 (0.333) \\ 
  lan\_namnVasternorrland & 0.346 (0.341) \\ 
  lan\_namnVastmanland & $-$3.807 (34,573.980) \\ 
  lan\_namnVastra gotalands lan & 0.020 (0.320) \\ 
    County\_specific\_mortality  & $-$0.00002 (0.007) \\ 
    \hline \\[-1.8ex] 
 \multicolumn{2}{l}{$^{*}$p$<$0.1; $^{**}$p$<$0.05; $^{***}$p$<$0.01} \\ 
    \end{tabular} 
\end{table}

\begin{table}[]
\centering
\footnotesize
\caption{All variables (* variable is included in factor analysis)}
\label{All_vars}
\begin{tabular}{lllll}
\hline
Variable & Description \\
\hline
&LOPNR & ID serial number \\ 
&group & Treatment group (group=1 for abiraterone) \\ 
& County & County council \\ 
&Diff\_time & Time between diagnosis and treatment \\
&SRE & Skeleton related event between diagnosis and treatment (SRE=1 if yes) \\
&Pain & Pain between diagnosis and treatment (Pain=1 if yes) \\
&   County\_specific\_mortality &   County specific mortality, the year of diagnosis  \\
 & SCORE\_D22 & Elixhouser index at diagnosis = 0 \\ 
  & SCORE\_D23 & Elixhouser index at diagnosis = 1-4 \\ 
   & SCORE\_D24 & Elixhouser index at diagnosis $>=5$ \\ 
 & d\_tot & Indicator of diabetes before treatment \\ 
 &o\_tot & Indicator of osteoporosis before treatment \\ 
 &c\_tot & Indicator of metastases before treatment \\ 
 &i48 & Indicator of atrial fibrillation and flutter before treatment \\ 
 &i21 & Indicator of acute myocardial infarction before treatment \\ 
 & h\_tot\_ov\_i & Indicator of other CVD before treatment \\ 
 &f\_tot & Indicator of fatigue before treatment \\ 
 &h\_tot & Indicator of any CVD before treatment \\ 
* & bfd\_1m & Number of visits, 1 month before diagnosis \\ 
* & bfd\_12m & Number of visits, 12 months before diagnosis \\ 
* & afd\_1m & Number of visits, 0-1 month after diagnosis \\ 
* & afd\_2m & Number of visits, 1-2 months after diagnosis \\ 
* & afd\_3m & Number of visits, 2-3 months after diagnosis \\ 
* & bft\_1m & Number of visits, 0-1 month before treatment\\ 
* & bft\_2m & Number of visits, 1-2 months before treatment \\ 
* & bft\_3m & Number of visits, 2-3 months before treatment \\ 
* & bft\_4m & Number of visits, 3-4 months before treatment \\ 
* & bft\_5m & Number of visits, 4-5 months before treatment \\ 
* & bft\_12m\_tot & Number of visits, 1 year before treatment \\ 
* & bft\_12m\_tot\_c619 & Number of C61.9 related visits, 1 year before treatment\\ 
* & msv\_days\_bd & Number of days in inpatient care before diagnosis \\ 
* & msv\_days\_c619 & Number of days in inpatient care related to C61.9 \\ 
* & msv\_tot\_prop\_c619 & Fraction of days in inpatient care related to C61.9 \\ 
* & msv\_days & Days in inpatient care between diagnosis and treatment \\ 
* & msv\_tot\_prop & Fraction of days in inpatient care between diagnosis and treatment \\ 
* & bfd\_60m & Number of visits, 5 years before diagnosis \\ 
* & nn & Visits per day between diagnosis and treatment \\
* & h\_tot\_bfd & Visits related to CVD before diagnosis \\ 
* & h\_tot\_bw & Visits related to CVD between diagnosis and treatment \\ 
* & d\_tot\_bfd & Visits related to diabetes before diagnosis \\ 
* & d\_tot\_bw & Visits related to diabetes between diagnosis and treatment \\ 
* & o\_tot\_bfd & Visits related to osteoporosis before diagnosis \\ 
* & o\_tot\_bw &  Visits related to osteoporosis between diagnosis and treatment \\ 
* & c\_tot\_bfd &  Visits related to metastases before diagnosis \\ 
* & c\_tot\_bw & Visits related to metastases between diagnosis and treatment \\ 
* & f\_tot\_bfd & Visits related to fatigue before diagnosis \\ 
* & f\_tot\_bw & Visits related to fatigue between diagnosis and treatment \\ 
* & i48\_bfd &  Visits related to atrial fibrillation and flutter before diagnosis \\ 
* & i21\_bfd & Visits related to acute myocardial infarction before diagnosis \\ 
* & h\_tot\_ov\_bfd & Visits related to other CVD before diagnosis \\
* & i48\_bw & Visits related to atrial fibrillation and flutter bw diagnosis and treatment \\ 
* & i21\_bw & Visits related to acute myocardial infarction bw diagnosis and treatment \\ 
* & h\_tot\_ov\_bw & Visits related to other CVD between diagnosis and treatment \\ 
\hline
\end{tabular}
\end{table}

\begin{table}[]
\centering
\ContinuedFloat
\footnotesize
\caption{All variables (* variable is included in factor analysis) cont.}
\label{All_vars}
\begin{tabular}{lll}
\hline
Variable & Description \\ 
\hline 
* & q1\_t & Inpatient care visits between diagnosis and treatment, quartile 1 \\ 
* & q2\_t & Inpatient care visits between diagnosis and treatment, quartile 2 \\ 
* & q3\_t & Inpatient care visits between diagnosis and treatment, quartile 3 \\ 
* & q4\_t & Inpatient care visits between diagnosis and treatment, quartile 4 \\ 
 &MEDS\_C08 & Number of prescriptions of ATC code C08 \\  &MEDS\_C07 & Number of prescriptions of ATC code C07 \\ 
 & MEDS\_A10 & Number of prescriptions of ATC code A10 \\ 
 & ALDER\_T & Age at treatment \\ 
  & UTBNFORGYMN & Educational level at  diagnosis: less than secondary school \\ 
 & UTBNGYMN & Educational level at diagnosis: secondary school \\ 
 & UTBNEFTERGYMN & Educational level at diagnosis: more than secondary school \\ 
 & Civil & Marital status at diagnosis (Civil=1 if partner) \\ 
& Fodelseland\_EU28 &Country of birth \\
* & LoneInk & Wage income at treatment \\ 
* & InkFNetto & Income from business at treatment \\ 
* & KapInk & Capital income at treatment \\ 	
* & DispInk & Disposable income at treatment \\ 
* & DispInkFam & Family disposable income at treatment \\ 
* & ForvErs & Earned income and work-related benefits at treatment \\ 
* & LoneInk\_1y & Wage income one year before treatment \\ 
* & InkFNetto\_1y & Income from business one year before treatment \\ 
* & KapInk\_1y & Capital income one year before treatment\\ 
*  & DispInk\_1y & Disposable income one year before treatment\\ 
* & DispInkFam\_1y & Family disposable income one year before treatment\\
* & ForvErs\_1y & Earned income and work-related benefits one year before treatment \\ 
* & LoneInk\_2y & Wage income two years before treatment \\ 
* & InkFNetto\_2y & Income from business two years before treatment \\ 
* & KapInk\_2y & Capital income two years before treatment\\ 
* & DispInk\_2y & Disposable two years before treatment\\ 
* & DispInkFam\_2y & Family disposable income two years before treatment\\
* & ForvErs\_2y & Earned income and work-related benefits two years before treatment \\ 
* & SjukRe & Sickness compensation at treatment \\ 
* & ArbLos & Unemployment benefits at treatment \\ 
* & ForTid & Early retirement benefit at treatment \\ 
* & SocInk & Social security benefits at treatment \\ 
* & SocBidrPersF & Social security benefits at treatment \\ 
* & SocBidrFam & Social security benefits of the family at treatment \\ 
* & SjukRe\_1y & Sickness compensation one year before treatment \\ 
* & ArbLos\_1y & Unemployment benefits one year before treatment \\ 
* & ForTid\_1y & Early retirement benefit one year before treatment \\ 
* & SocInk\_1y & Social security benefits one year before treatment \\
* & SocBidrPersF\_1y & Social security benefits one year before treatment \\ 
* & SocBidrFam\_1y & Social security benefits of the family one year before treatment \\ 
* & SjukRe\_2y & Sickness compensation two years before treatment \\ 
* & ArbLos\_2y & Unemployment benefits two years before treatment \\ 
* & ForTid\_2y & Early retirement benefit two years before treatment \\ 
* & SocInk\_2y & Social security benefits two years before treatment \\ 
* & SocBidrPersF\_2y & Social security benefits two years before treatment \\ 
* & SocBidrFam\_2y & Social security benefits of the family two years before treatment \\ 
* & AldPens & Old-age pensions at treatment \\ 
* & SumTjP & Occupational pensions one year before treatment \\ 
* & PrivPens & Private pensions two years before treatment \\ 
* & AldPens\_1y & Old-age pensions at treatment \\ 
* & SumTjP\_1y & Occupational pensions one year before treatment\\ 
* & PrivPens\_1y & Private pensions two years before treatment \\ 
* & AldPens\_2y & Old-age pensions at treatment \\ 
* & SumTjP\_2y & Occupational pensions one year before treatment\\ 
* & PrivPens\_2y & Private pensions two years before treatment \\ 
\hline
\end{tabular}
\end{table}

\begin{table}[]
\centering
\ContinuedFloat
\footnotesize
\caption{All variables (* variable is included in factor analysis) cont.}
\label{All_vars}
\begin{tabular}{lll}
\hline
Variable & Description \\ 
\hline 
* & LoneInk\_D & Wage income at diagnosis \\ 
* & InkFNetto\_D & Income from business at diagnosis \\ 
* & KapInk\_D & Capital income at diagnosis \\ 	
* & DispInk\_D & Disposable income at diagnosis \\ 
* & DispInkFam\_D & Family disposable income at diagnosis \\ 
* & LoneInk\_1y\_D & Wage income one year before diagnosis \\ 
* & InkFNetto\_1y\_D & Income from business one year before diagnosis \\ 
* & KapInk\_1y\_D & Capital income one year before diagnosis\\ 
*  & DispInk\_1y\_D & Disposable income one year before diagnosis\\ 
* & DispInkFam\_1y\_D & Family disposable income one year before diagnosis\\
* & LoneInk\_2y\_D & Wage income two years before diagnosis \\ 
* & InkFNetto\_2y\_D & Income from business two years before diagnosis \\ 
* & KapInk\_2y\_D & Capital income two years before diagnosis\\ 
* & DispInk\_2y\_D & Disposable two years before diagnosis\\ 
* & DispInkFam\_2y\_D & Family disposable income two years before diagnosis\\
* & SjukRe\_D & Sickness compensation at diagnosis \\ 
* & ArbLos\_D & Unemployment benefits at diagnosis \\ 
* & ForTid\_D & Early retirement benefit at diagnosis \\ 
* & SocInk\_D & Social security benefits at diagnosis \\ 
* & SocBidrPersF\_D & Social security benefits at diagnosis \\ 
* & SocBidrFam\_D & Social security benefits of the family at diagnosis \\ 
* & SjukRe\_1y\_D & Sickness compensation one year before diagnosis \\ 
* & ArbLos\_1y\_D & Unemployment benefits one year before diagnosis \\ 
* & ForTid\_1y\_D & Early retirement benefit one year before diagnosis \\ 
* & SocInk\_1y\_D & Social security benefits one year before diagnosis \\
* & SocBidrPersF\_1y\_D & Social security benefits one year before diagnosis \\ 
* & SocBidrFam\_1y\_D & Social security benefits of the family one year before diagnosis \\ 
* & SjukRe\_2y\_D & Sickness compensation two years before diagnosis \\ 
* & ArbLos\_2y\_D & Unemployment benefits two years before diagnosis \\ 
* & ForTid\_2y\_D & Early retirement benefit two years before diagnosis \\ 
* & SocInk\_2y\_D & Social security benefits two years before diagnosis \\ 
* & SocBidrPersF\_2y\_D & Social security benefits two years before diagnosis \\ 
* & SocBidrFam\_2y\_D & Social security benefits of the family two years before diagnosis \\ 
* & AldPens\_D & Old-age pensions at diagnosis \\ 
* & SumTjP\_D & Occupational pensions one year before diagnosis \\ 
* & PrivPens\_D & Private pensions two years before diagnosis \\ 
* & AldPens\_1y\_D & Old-age pensions at diagnosis \\ 
* & SumTjP\_1y\_D & Occupational pensions one year before diagnosis\\ 
* & PrivPens\_1y\_D & Private pensions two years before diagnosis \\ 
* & AldPens\_2y\_D & Old-age pensions at diagnosis \\ 
* & SumTjP\_2y\_D & Occupational pensions one year before diagnosis\\ 
* & PrivPens\_2y\_D & Private pensions two years before diagnosis \\ 
\hline
\end{tabular}
\end{table}

\end{document}